\documentclass[ 
reprint,
superscriptaddress,
longbibliography,
 amsmath,amssymb,
 aps,
]{revtex4-2}

\usepackage{graphicx}% Include figure files
\usepackage{dcolumn}% Align table columns on decimal point
\usepackage{bm}% bold math
\usepackage{float}
\usepackage[normalem]{ulem}
\usepackage{microtype}
\usepackage[dvipsnames]{xcolor}
\usepackage[breaklinks]{hyperref}% add hypertext capabilities
\usepackage[separate-uncertainty = true,multi-part-units=single]{siunitx}
\DeclareSIUnit\sq{sq}
\DeclareSIUnit\T{T}
\DeclareSIUnit\dBm{dBm}

\usepackage{subfiles}

\definecolor{edumur}{RGB}{46,116,181}

\usepackage[normalem]{ulem}

\usepackage{lineno}
\begin{document}

%\preprint{APS/123-QED}

%\title{Phase and gate control of the harmonic's content of the current phase relation in a germanium based Josephson field effect transistor.}

\title{Gate- and flux-tunable sin(2$\varphi$) Josephson element with proximitized Ge-based junctions}

\author{Axel Leblanc}
\email{E-mail: axel.leblanc@cea.fr}
\affiliation{Univ. Grenoble Alpes, CEA, Grenoble INP, IRIG, PHELIQS, 38000 Grenoble, France}
\author{Chotivut Tangchingchai}
\affiliation{Univ. Grenoble Alpes, CEA, Grenoble INP, IRIG, PHELIQS, 38000 Grenoble, France}
\author{Zahra Sadre Momtaz}
\affiliation{Institut Néel, CNRS/UGA, Grenoble 38042, France}
\author{Elyjah Kiyooka}
\affiliation{Univ. Grenoble Alpes, CEA, Grenoble INP, IRIG, PHELIQS, 38000 Grenoble, France}
\author{Jean-Michel Hartmann}
\affiliation{Univ. Grenoble Alpes, CEA, LETI, 38000 Grenoble, France}
\author{Fr\'ed\'eric Gustavo}
\affiliation{Univ. Grenoble Alpes, CEA, Grenoble INP, IRIG, PHELIQS, 38000 Grenoble, France}
\author{Jean-Luc Thomassin}
\affiliation{Univ. Grenoble Alpes, CEA, Grenoble INP, IRIG, PHELIQS, 38000 Grenoble, France}
\author{Boris Brun}
\affiliation{Univ. Grenoble Alpes, CEA, Grenoble INP, IRIG, PHELIQS, 38000 Grenoble, France}
\author{Vivien Schmitt}
\affiliation{Univ. Grenoble Alpes, CEA, Grenoble INP, IRIG, PHELIQS, 38000 Grenoble, France}
\author{Simon Zihlmann}
\affiliation{Univ. Grenoble Alpes, CEA, Grenoble INP, IRIG, PHELIQS, 38000 Grenoble, France}
\author{Romain Maurand}
\affiliation{Univ. Grenoble Alpes, CEA, Grenoble INP, IRIG, PHELIQS, 38000 Grenoble, France}
\author{\'Etienne Dumur}
\affiliation{Univ. Grenoble Alpes, CEA, Grenoble INP, IRIG, PHELIQS, 38000 Grenoble, France}
\author{Silvano De Franceschi}
\email{E-mail: silvano.defranceschi@cea.fr}
\affiliation{Univ. Grenoble Alpes, CEA, Grenoble INP, IRIG, PHELIQS, 38000 Grenoble, France}
\author{François Lefloch}
\email{E-mail: francois.lefloch@cea.fr}
\affiliation{Univ. Grenoble Alpes, CEA, Grenoble INP, IRIG, PHELIQS, 38000 Grenoble, France}

\date{\today}% It is always \today, today,

\begin{abstract}

%Josephson junctions (JJs) made out of hybrid superconductor(S) and semiconductor(Sm) materials distinguish themselves from tunnel JJs by gate tunability and multi-harmonic current phase relation (CPR). These two ingredients are essential for the realization of a $\sin(2\varphi)$ Josephson element in an inductance-free superconducting quantum-interference device (SQUID), which is the most challenging building block of hybrid parity-protected superconducting qubits. This work studies the CPR of a SQUID embedding Josephson field-effect transistors and reports an experimental direct measurement of a $\pi$-periodic CPR also demonstrating its fine gate and flux tunability. Notably, a quantitative analysis including a complete modeling of the circuits allows to establish that the charge-4e supercurrent contributes up to \SI{95.2}{\percent} of the total supercurrent flowing through the device.

Hybrid superconductor-semiconductor Josephson field-effect transistors (JoFETs) function as Josephson junctions with a gate-tunable critical current. Additionally, they can feature a non-sinusoidal current-phase relation (CPR) containing multiple harmonics of the superconducting phase difference, a so-far underutilized property. In this work, we exploit this multi-harmonicity to create a Josephson circuit element with an almost perfectly $\pi$-periodic CPR, indicative of a largely dominant charge-4e supercurrent transport. Such a Josephson element was recently proposed as the basic building block of a protected superconducting qubit. Here, it is realized using a superconducting quantum interference device (SQUID) with low-inductance aluminum arms and two nominally identical JoFETs. The latter are fabricated from a SiGe/Ge/SiGe quantum-well heterostructure embedding a high-mobility two-dimensional hole gas. By carefully adjusting the JoFET gate voltages and finely tuning the magnetic flux through the SQUID close to half a flux quantum, we achieve a regime where the $\sin(2\varphi)$ component accounts for more than \SI{95}{\percent} of the total supercurrent. This result demonstrates a new promising route for the realization of superconducting qubits with enhanced coherence properties.

%Here we report an experimental study of the CPR harmonic content of a SQUID embedding Josephson field-effect transistors. The SQUID is gate-balanced and frustrated at half flux quantum to demonstrate the realization of a $\pi$-periodic Josephson element. Notably, the charge-4e supercurrent contributes up to \SI{96}{\percent} of the total supercurrent flowing through the device.

\end{abstract}

\maketitle

\section{Introduction}

Quantum information processing requires qubits with long coherence time enabling high fidelity quantum gates. Over the past two decades, superconducting circuits have led to the realization of quantum processors of ever-growing size made of qubits with steadily improving fidelities \cite{Bravyi_IBM2022}. This way, superconducting qubits have become one of the most advanced physical platforms for quantum computing. Progress has been driven by material engineering and optimization, as well as by the development of new device concepts capable of providing a growing level of protection against noise sources in the environment \cite{gyenis_moving_2021, calzona_multi-mode_2022, danon_protected_2021}. Qubit protection against relaxation and dephasing processes can be granted from the symmetry properties of the qubit Hamiltonian. In this direction, a variety of possible solutions have been proposed and only partly explored  \cite{doucot_pairing_2002, kitaev_protected_2006, koch_charge-insensitive_2007, manucharyan_fluxonium_2009, devoret_superconducting_2013, kalashnikov_bifluxon_2020, gyenis_experimental_2021}. One of the leading ideas is to create superconducting qubits whose two lowest energy states are associated with odd and even numbers of Cooper pairs in a superconducting island, respectively. Due to the different parity, these states are orthogonal to each other in both charge and phase space \cite{ioffe_possible_2002, gladchenko_superconducting_2009, bell_protected_2014, smith_superconducting_2020, larsen_parity-protected_2020, schrade_protected_2022, maiani_entangling_2022, smith_magnifying_2022}. This type of parity-protected qubit requires a parity-preserving Josephson element that only allows the coherent transfer of correlated pairs of Cooper pairs, which translates into devising a Josephson circuit with a $\pi$-periodic, $\sin(2\varphi)$ current phase relation (CPR). 

Some proposals to engineer such a $\sin(2\varphi)$ qubits rely on conventional $\sin(\varphi)$ Josephson junctions, either arranged into large arrays \cite{bell_protected_2014} or embedded in a superconducting quantum interference device (SQUID) together with extremely large inductances \cite{smith_superconducting_2020}. The practical realisation of these ideas is technologically challenging and some significant experimental  progress was reported only recently \cite{smith_magnifying_2022}.
Another approach is to leverage the multi-harmonic  CPR and the gate tunability of superconductor(S)-semiconductor(Sm) Josephson field-effect transistors (JoFETs)\cite{spanton_currentphase_2017, english_observation_2016, golubov_current-phase_2004, beenakker_quantum_1992, ueda_evidence_2020, nichele_relating_2020, nanda_current-phase_2017, sochnikov_nonsinusoidal_2015, portoles_tunable_2022, baumgartner_supercurrent_2022}. Various signatures of $\sin(2\varphi)$ Josephson elements were recently reported \cite{valentini_parity-conserving_2024, ciaccia_charge-4e_2024, leblanc_nonreciprocal_2023, banszerus_voltage-controlled_2024, de_lange_realization_2015, messelot_direct_2024} and harnessed to demonstrate some first experimental evidence of parity protection \cite{larsen_parity-protected_2020}.
However, a direct measurement of a $\sin(2\varphi)$ CPR and precise quantitative evaluation of its harmonic purity and its tunability have been missing so far. These important aspects are addressed in the present work.   
Our experimental study takes advantage of a recently developed S-Sm platform based on SiGe/Ge/SiGe quantum-well heterostructures. We investigate the CPR of a SQUID embedding two gate-tunable Josephson junctions, in short a G-SQUID. We demonstrate ample gate and magnetic-flux control of the Josephson harmonic content. 
%and, in particular, the ratio between the $\sin(2\varphi)$ (i.e. charge-4e) and $\sin(\varphi)$ (i.e. charge-2e)  contributions  to the supercurrent. 
In particular, our quantitative analysis based on a fully comprehensive model of our circuit, reveals that the desired $\sin(2\varphi)$ (i.e. charge-4e)  contribution to supercurrent can reach up to \SI{95.2}{\percent} of the total supercurrent at half flux quantum through the SQUID.
This achievement is a significant step forward in the development and optimization of a semiconductor-based parity-protected qubit.

\section{Device and measurement setup}

\begin{figure*}
\includegraphics[width=1\textwidth]{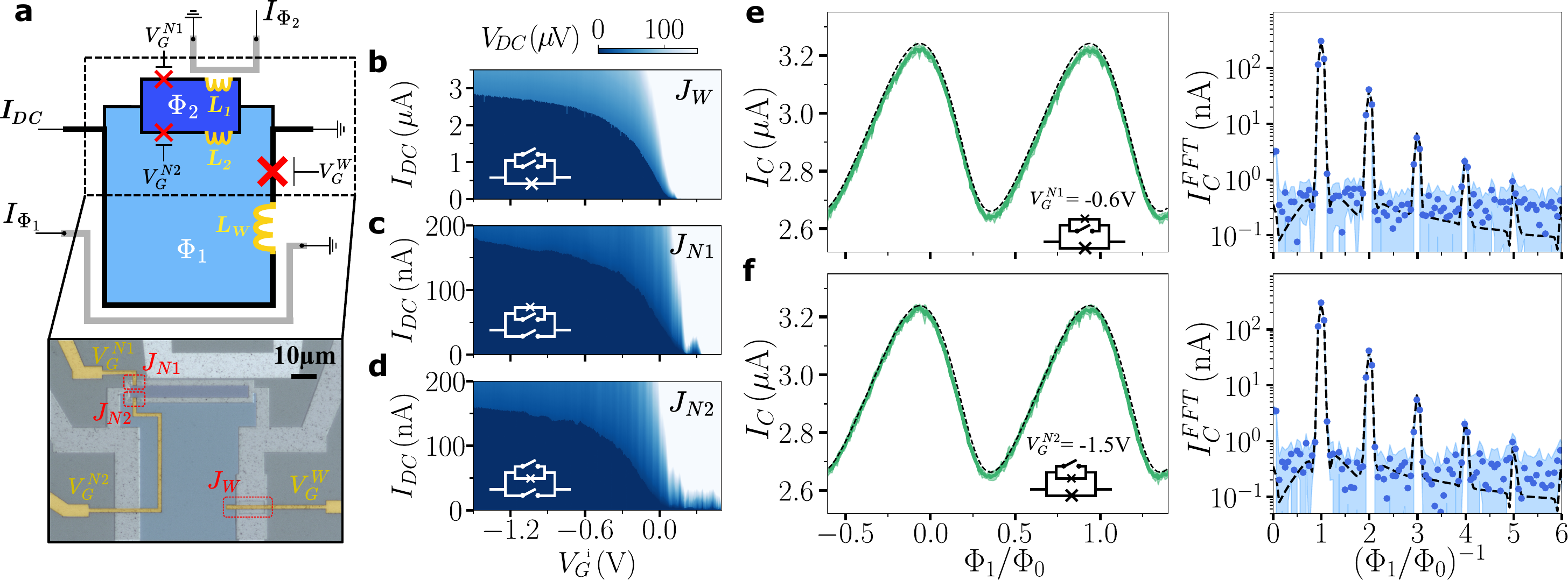}
\caption{\textbf{Double SQUID device and single-JoFET characteristics.}  \textbf{a}, Schematic and scanning electron micrograph of the device. A Ge-based SQUID (G-SQUID), embedding JoFETs $J_\mathrm{N1}$ and $J_\mathrm{N2}$, is connected in parallel to a wider JoFET ($J_\mathrm{W}$) used as a reference Josephson junction for current-phase-relation (CPR) measurements.
The  aluminum arms are modeled by three inductances: $L_1$, $L_2$ and $L_\mathrm{W}$. \textbf{b},\textbf{c},\textbf{d}, Current-biased  measurements of the JoFET characteristics for  $J_\mathrm{W}$, $J_\mathrm{N1}$, and $J_\mathrm{N2}$, respectively. In each panel, the measured source-drain voltage is plotted as a function of gate voltage, $V_\mathrm{G}^i$ ($i=$ W,N1,N2), and source-drain current bias $I_{DC}$.
\textbf{e} (resp. \textbf{f}), Left: critical current, $I_C$, as a function of magnetic  flux $\Phi_1$ through the large loop in \textbf{a}. The reference JoFET $J_\mathrm{W}$ is biased to strong accumulation ($V_\mathrm{G}^W = \SI{-1.5}{\volt}$), $J_\mathrm{N2}$ (resp. $J_\mathrm{N1}$) is pinched off and $V_\mathrm{G}^\mathrm{N1}=\SI{-0.6}{\volt}$ (resp. $V_\mathrm{G}^\mathrm{N2}=\SI{-1.5}{\volt}$).
The $I_C$ oscillations are a direct  measurement of $J_\mathrm{N2}$ (resp. $J_\mathrm{N1}$) CPR. Right: Fast Fourier transform (FFT) of the CPR on the left, calculated over 15 $\Phi_0$.
Dashed line: calculated FFT based on the circuit model in \textbf{a} with parameters obtained from a fit of the data in Fig.~2a.}
\label{fig:characterization}
\end{figure*}

The G-SQUID (shown in Fig.~\ref{fig:characterization}a) consists of an aluminum superconducting loop with nearly symmetric arms embedding two nominally identical JoFETs fabricated out of a SiGe/Ge/SiGe quantum-well heterostructure. The compressively strained Ge quantum well lies \SI{22}{\nm} beneath the semiconductor surface and hosts a two-dimensional hole gas exhibiting a mobility of $10^5$ cm$^2$/Vs measured at a carrier density of 
$6 \times 10^{11}$ cm$^{-2}$  .
The two JoFETs, $J_\mathrm{N1}$ and $J_\mathrm{N2}$, have a \SI{1}{\um}-wide and a  \SI{300}{\nm}-long Ge channel (for more details on the JoFETs see Supp.~S1).
The G-SQUID is embedded in a second larger loop together with a wider, reference JoFET, $J_\mathrm{W}$, (\SI{10}{\um}-wide and \SI{300}{\nm}-long Ge channel),  
enabling a direct CPR measurement     \cite{della_rocca_measurement_2007, golubov_current-phase_2004, spanton_currentphase_2017, murani_ballistic_2017, stoutimore_second-harmonic_2018, endres_currentphase_2023, frolov_current-phase_2005, ginzburg_determination_2018}. To this purpose, $J_\mathrm{W}$ is designed to have a critical current much larger than those of $J_\mathrm{N1}$ and $J_\mathrm{N2}$.
The small and large superconducting loops are locally flux biased by means of two \SI{10}{\um}-wide and \SI{50}{\nm}-thick Al lines  whose cross-talks have been calibrated {\it in-situ} and then implicitly  compensated throughout the rest of the paper (see Suppl. S6).

Furthermore, we stress that the Al arms have small but non-negligible inductances, mostly of kinetic origin, that we label as $L_1$, $L_2$ and $L_\mathrm{W}$.
As our later analysis will reveal, properly extracting the intrinsic harmonic content of the JoFET CPRs from the measurements requires taking these inductances into account. 
In the rest of the paper, all of the calculated curves are obtained using the circuit model shown in Fig.~\ref{fig:characterization}a (see Suppl. S2 for more details).

All measurements were performed in a dilution refrigerator at  a base  temperature  of \SI{38}{\milli\kelvin}. The fabrication process and the measurement methods are very similar to those discussed in \cite{leblanc_nonreciprocal_2023, hartmann_epitaxy_2023}.

\section{characterisation of the individual JOFETs}

To access the individual DC transport characteristic of a given JoFET in such a parallel configuration, we purposely apply large positive gate voltages ($\approx \SI{1.5}{\volt}$) to the other JoFETs, thereby suppressing current flow  through their respective arms.
The resulting individual  characteristics of $J_\mathrm{W}$, $J_\mathrm{N1}$ and $J_\mathrm{N2}$ as a function of their respective gate voltages are shown on color scale in Figs.~\ref{fig:characterization}b-d, with the corresponding circuit measurement schematics displayed in the insets.

For all the JoFETs, the current at which the device switches from superconducting to normal state, close to the critical current  $I_\mathrm{C}$, is clearly visible as an abrupt change of the measured source-drain voltage drop from $0$ to a finite value.  The three JoFETs exhibit a similar behavior denoting consistent properties of the Ge channel and the superconducting contacts. In particular, 
we note that the two narrow JoFETs, designed to be identical, have very similar $I_{C}^\mathrm{Ni}(V_\mathrm{G}^\mathrm{Ni})$ characteristics.

By varying the magnetic flux, $\Phi_1$, through the larger loop we can sequentially measure the CPR of each  JoFET in the G-SQUID. To this aim,  the reference JoFET, $J_W$, is biased at full accumulation ($V_\mathrm{G}^\mathrm{W}=\SI{-1.5}{\volt}$) such that the necessary condition $I_{C}^\mathrm{W} \gg I_{C}^\mathrm{N_1},I_{C}^\mathrm{N_2}$ is fulfilled \cite{della_rocca_measurement_2007,babich_limitations_2023}. To measure the $J_{N_1}$ ($J_{N_2}$) CPR, we apply $V_\mathrm{G}^\mathrm{N_1}=\SI{-0.6}{\volt}$ 
($V_\mathrm{G}^\mathrm{N2} = \SI{-1.5}{\volt}$) while $J_{N_2}$ ($J_{N_1}$) is pinched off by setting 
$V_\mathrm{G}^\mathrm{N2} = \SI{1.5}{\volt}$ ($V_\mathrm{G}^\mathrm{N1} = \SI{1.5}{\volt}$). The on-state voltages $V_\mathrm{G}^\mathrm{N1} = \SI{-0.6}{\volt}$ and $V_\mathrm{G}^\mathrm{N2} = \SI{-1.5}{\volt}$ are chosen to obtain equal amplitudes of the first CPR harmonic. As we shall see below, operating the G-SQUID at these gate voltages enables the suppression of the first harmonic by flux-induced destructive interference, hence leaving a dominant sin$(2\varphi)$ component.   

The measured CPRs are shown in Figs.~\ref{fig:characterization}e,f together with the respective Fourier transforms. A total of 15 flux periods were measured in order to ensure sufficient resolution of the harmonics in Fourier space. Each reported $I_\mathrm{C}$ data point represents the median value obtained from 10 measurements, with the light green area indicating $\pm~1$ standard deviation. Both CPRs are clearly skewed and we distinguish up to five  harmonics. This multi-harmonicity indicates a high transparency of the superconducting contacts, which is consistent with earlier observations with similar devices \cite{leblanc_nonreciprocal_2023, valentini_parity-conserving_2024}.  Yet, as we shall discuss below, the higher harmonics, especially the fourth and fifth one, have largely enhanced amplitudes due to the finite inductance of the aluminum arms. 

%We now explore the magnetic flux dependence of the two narrow JoFETs by measuring their CPR, shown in Fig.~\ref{fig:characterization}e,f alongside their corresponding Fourier transform, again using the pinched-off method to ensure single JoFET characterization (see schematics).
%In such measurement, the wide JoFET is used as a ``reference'' JoFET  with much larger current at full accumulation, $V_\mathrm{G}^\mathrm{W}=\SI{-1.5}{\volt}$.
%The two JoFET gates voltages were tuned to equalize the first CPR harmonic in order of symmetrizing the G-SQUID ending up with $V_\mathrm{G}^\mathrm{N1} = \SI{-1.5}{\volt}$ and $V_\mathrm{G}^\mathrm{N2} = \SI{-0.6}{\volt}$.

%The Fourier transforms in the right panels of Figs.~\ref{fig:characterization}e,f are calculated on a flux interval 
%In total, 15 flux periods were measured to ensure sufficient resolution of the harmonics in Fourier space with each reported $I_\mathrm{C}$ data point being the median value over 10 measurements with the light green area corresponds to $\pm~1$ standard deviation.

%These measurements show that the flux response is skewed and composed of up to 5 distinguishable harmonics as a consequence of the JoFET's high transparency, confirming the multi-harmonicity already observed on similar devices\cite{leblanc_nonreciprocal_2023, valentini_parity-conserving_2024}.
 
%The experimental demonstration of multi-harmonic JoFET being fulfilled,  we now focus on the G-SQUID response and show how to engineer a sin(2$\varphi$) Josephson element.

\section{sin(2$\varphi$) Josephson element}

%\begin{figure*}
%\includegraphics[width=1\textwidth]{Fig/fig2_V5.pdf}
%\caption{\textbf{Frequency doubling of the G-SQUID CPR.}  The G-SQUID is symmetrized by equalizing the amplitudes  ($V_\mathrm{G}^\mathrm{N1}=\SI{-1.5}{\volt}$, $V_\mathrm{G}^\mathrm{N2}=\SI{-0.6}{\volt}$) and $J_\mathrm{W}$ is kept in accumulation ($V_\mathrm{W}=\SI{-1.5}{\volt}$). \textbf{a}, Critical current as a function of the two compensated flux bias $\Phi_1$ and $\Phi_2$. The oscillations are associated to the $\mathrm{G-SQUID}$ CPR. \textbf{b,c,d} $I_\mathrm{C}$ as a function of $\Phi_1$ for $\Phi_2$ = 0, $\Phi_0/4$ and $\Phi_0/2$. As shown by the Fourier transforms [\textbf{e,f,g}], in the first case all harmonics are present, in the second case even harmonics are reduced and in the third case odd harmonics are reduced leading to the doubling of the CPR frequency.
%In black dashed line, we show the value of the G-SQUID CPRs and their Fourier transform from our circuit model.}
%\label{fig:data}
%\end{figure*}

\begin{figure*}
\includegraphics[width=1\textwidth]{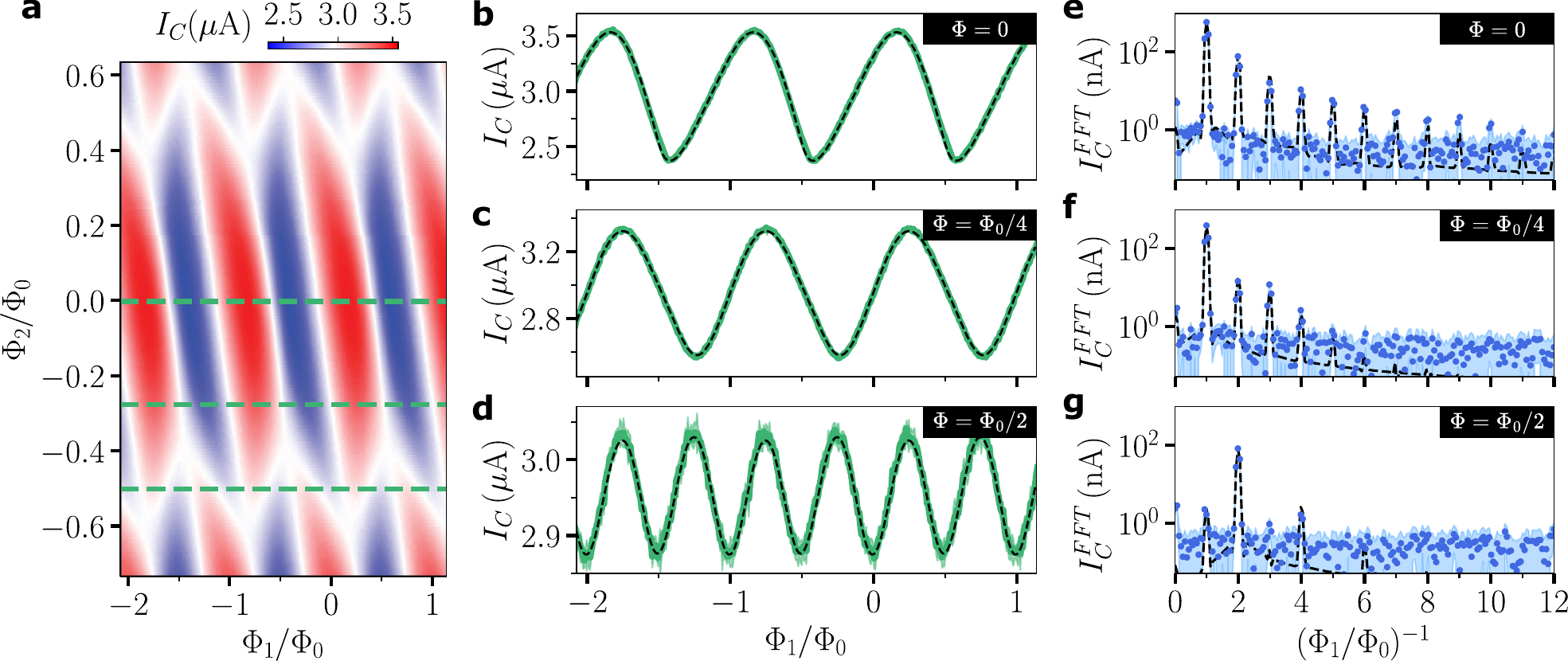}
\caption{\textbf{Frequency doubling of the G-SQUID CPR.}  The G-SQUID is symmetrized by setting $V_\mathrm{G}^\mathrm{N1}=\SI{-1.5}{\volt}$ and  $V_\mathrm{G}^\mathrm{N2}=\SI{-0.6}{\volt}$, which equalizes the amplitudes of the first harmonics in $J_\mathrm{N1}$ and $J_\mathrm{N2}$. $J_\mathrm{W}$ is kept in strong accumulation ($V_\mathrm{W}=\SI{-1.5}{\volt}$). \textbf{a}, Critical current as a function of the two compensated magnetic fluxes, $\Phi_1$ and $\Phi_2$, threading the large superconducting loop and the G-SQUID, respectively.  \textbf{b,c,d}, G-SQUID CPR ($I_\mathrm{C}$ vs $\Phi_1$) for $\Phi_2$ = 0, $\Phi_0/4$ and $\Phi_0/2$, i.e. at the line cuts denoted by green lines in \textbf{a}. \textbf{e,f,g}, the FFTs obtained from the CPRs in \textbf{b},\textbf{c}, and \textbf{d}, respectively, are calculated over 15 $\Phi_0$. Dashed lines in \textbf{b,c,d} (\textbf{e,f,g}) are  calculated CPRs  (FFTs) based on the circuit model in Fig.~1a with parameters obtained from a fit of the data in \textbf{a}. In \textbf{d}, following a suppression of the odd harmonics (clearly shown in \textbf{d}), we observe the doubling of the CPR frequency as expected for a $\sin(2\varphi)$ Josephson element.}
\label{fig:data}
\end{figure*}

With the two JoFETs $J_\mathrm{N1}$ and $J_\mathrm{N2}$ independently characterized, we now  turn to the study of the G-SQUID CPR, once again using $J_W$ as a reference. With the ultimate goal to engineer a sin(2$\varphi$) Josephson element, we symmetrize the G-SQUID by applying $V_\mathrm{G}^\mathrm{N1} = \SI{-0.6}{\volt}$ and $V_\mathrm{G}^\mathrm{N2} = \SI{-1.5}{\volt}$.

%We remind that the wide JoFET $J_W$ is used as a reference JoFET while the two others, forming the G-SQUID, are kept in the symmetric regime described previously.

Figure ~\ref{fig:data}a shows a measurement of the G-SQUID critical current as a function of $\Phi_1$ and $\Phi_2$, the latter being the magnetic flux through the G-SQUID loop. This is the most important data set. 
%effectively revealing the flux tunability of the G-SQUID CPR.
% The $I_\mathrm{C}(\Phi_1)$ oscillations are associated with the SQUID CPR.
%Remarkably the G-SQUID CPR linear phase evolution with respect to $\Phi_2$ appeared in such a map as a tilted pattern, see Supp.~S3 for deeper analysis.
We fit the entire two-dimensional plot to the circuit model shown in Fig.~\ref{fig:characterization}a, with fixed inductances of the Al arms (see Supp.~S1) and twelve free parameters accounting for the amplitudes of the first four harmonics of the three JoFETs (see Supp.~S2 for details).  Interestingly, the fit yields negligible amplitudes for all the fourth-order harmonics, implying that only three harmonics per JoFET are sufficient to reproduce the data. 
This outcome apparently contrasts with the experimental data in Figs.~\ref{fig:characterization}e,f, where up to five harmonics can be distinguished for both  $J_\mathrm{N1}$ and $J_\mathrm{N2}$. 
%In fact, higher harmonics can emerge due to the finite inductance of the aluminum arms. 
The discrepancy arises from the finite inductances of the Al arms \cite{ginzburg_determination_2018, lecocq_dynamique_2011}. We estimate $L_1 \approx L_2 \approx \SI{50}{\pico\henry}$ (i.e. 20 times smaller than the inductances of $J_\mathrm{N1}$ and $J_\mathrm{N2}$) and $L_W \approx \SI{210}{\pico\henry}$. Even such relatively small inductances can significantly enhance the harmonic amplitudes. The enhancement becomes  proportionally larger as the harmonic order increases \cite{ginzburg_determination_2018, lecocq_dynamique_2011}. This aspect is fully captured by our circuit model. Indeed, the dashed lines in 
Figs.~\ref{fig:characterization}e,f are  calculated using the parameters extracted from the fit of Fig.~\ref{fig:data}a. The experimental data are accurately reproduced despite the fact that only three harmonics effectively contribute to the CPR of each JoFET.

In Figs.~\ref{fig:characterization}e,f  the arm inductances amplify the $\mathrm{1^{st}}$, $\mathrm{2^\mathrm{nd}}$, and $\mathrm{3^{rd}}$ harmonics of $J_\mathrm{N1}$ ($J_\mathrm{N2}$) by 7(6)\%, 22(30)\%, and 250(244)\%, respectively, and lead to the emergence of a $\mathrm{4^{th}}$ and a $\mathrm{5^{th}}$ harmonic \cite{footnote}. 
Hence our analysis reveals the importance of including even small contributions of arm inductances in order to avoid a crude overestimation of the harmonic amplitudes.
Finally, we remark that arm inductances can also induce a phase shift in the CPR \cite{lecocq_dynamique_2011}.
We finally address the magnetic-flux dependence of the G-SQUID CPR.
At $\Phi_2=0$ (Figs. 2b,e), we expect the G-SQUID CPR to be the sum of the $J_\mathrm{N1}$ and $J_\mathrm{N2}$ CPRs shown in Figs. 1e,f. Instead, we observe a CPR containing about ten harmonics. Moreover, all harmonics beyond the first one 
exhibit amplitudes clearly larger than expected from a simple addition. Fully captured by our circuit model (see dashed lines in Figs. 2b,e), this finding is mostly a consequence of the moderate ratio between the critical current of the reference junction $J_\mathrm{W}$ and the one of the G-SQUID \cite{babich_limitations_2023}, i.e.  $I^\mathrm{W}_\mathrm{C}/I^\mathrm{G-SQUID}_\mathrm{C} \approx 5$ at $\Phi_2=0$.

At $\Phi_2=\Phi_0/4$ (Fig 2c,f), $J_\mathrm{N1}$ and $J_\mathrm{N2}$ are dephased by $\pi/2$, resulting in a destructive interference between even harmonics.
The 2\textsuperscript{nd} and 4\textsuperscript{th}  harmonics are consequently suppressed while the 1\textsuperscript{st} and 3\textsuperscript{rd}  preserve the same amplitude. The resulting CPR is clearly less skewed than at $\Phi_2=0$.
From our model, we conclude that the residual  2\textsuperscript{nd} and 4\textsuperscript{th} harmonics are again due to a moderate ratio  $I^\mathrm{W}_\mathrm{C}/I^\mathrm{SQUID}_\mathrm{C}$.
Increasing the $J_\mathrm{W}$ critical current by a factor  ten would further suppress the 2\textsuperscript{nd} harmonic by the same factor.

At $\Phi_2=\Phi_0/2$ (Fig 2d,g), a $\pi$ phase shift induces a destructive interference between the odd harmonics of $J_\mathrm{N1}$ and $J_\mathrm{N2}$ with reduction of the G-SQUID critical current. 
Following the suppression of the 1\textsuperscript{st} and 3\textsuperscript{rd} harmonics, the 2\textsuperscript{nd} harmonic becomes the dominant one resulting in the emergence of a $\Phi_0/2$ flux periodicity in the CPR. 
In conclusion,  at half flux quantum, the G-SQUID behaves as a $\sin(2\varphi)$ Josephson element.

\section{Harmonic tuning and sin(2$\varphi$) purity}

\begin{figure}
\vspace{0.5cm}
\includegraphics[width=0.45\textwidth]{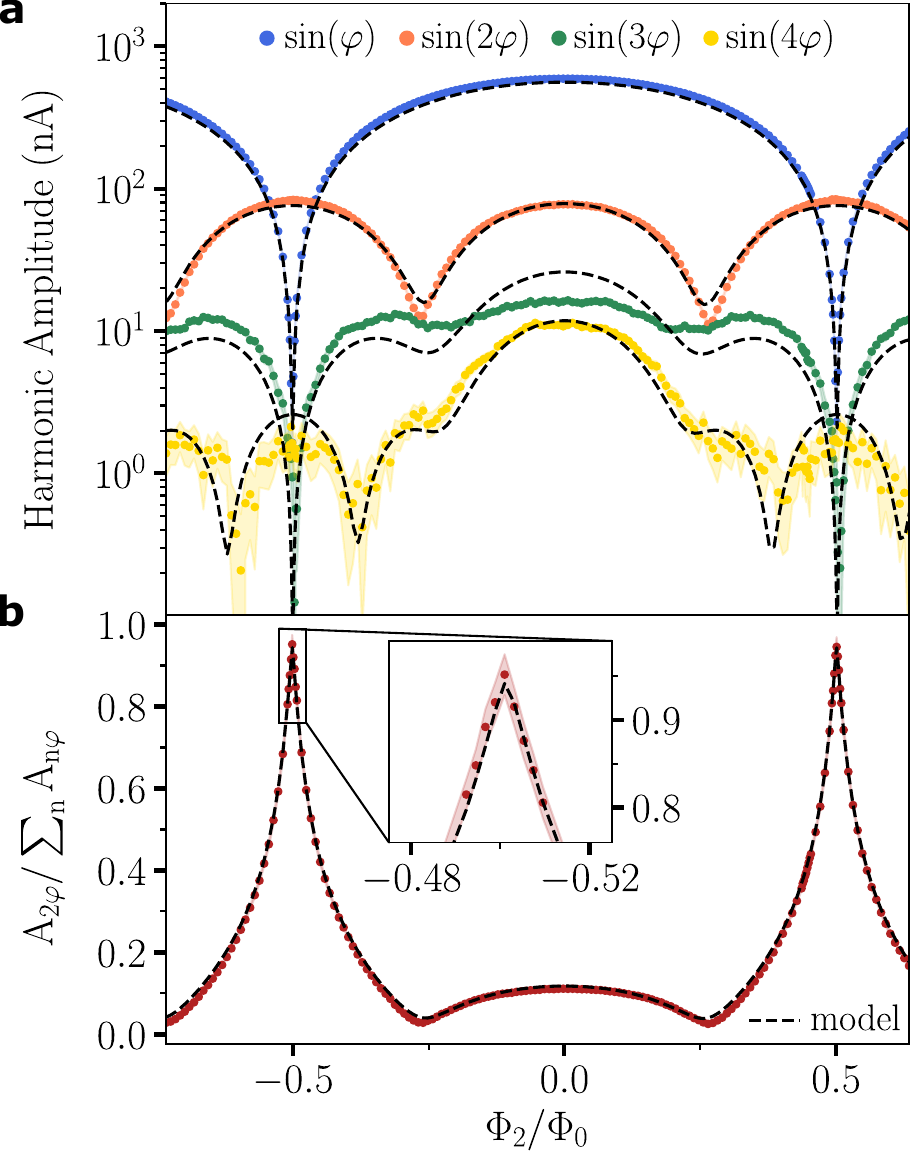}
\caption{\textbf{Flux modulation of the G-SQUID harmonics}  \textbf{a}, Amplitudes of the first four harmonics in the G-SQUID CPR as a function of $\Phi_2$. The first harmonic vanishes at $\Phi_0/2$,  while the second one vanishes at $\Phi_0/4$. \textbf{b}, Flux dependence of  $\sin(2\varphi)$ purity, which is define as the ratio between the amplitude of the second harmonic, $A_{2\varphi}$, and the sum of the all four harmonic amplitudes, $\Sigma_n A_{n\varphi}$. The $\sin(2\varphi)$ purity has a sharp maximum at $\Phi_0/2$ where it reaches \SI{95.2(2.4)}{\percent}. Inset: close-up around the maximum. 
Colored bands in \textbf{a} and \textbf{b} represent the $\pm \sigma$ standard deviation originating from the experimental uncertainty on the CPR data points.  Black dashed lines in \textbf{a} and \textbf{b} represent the harmonic amplitudes calculated from our circuit model using the fit parameters from of Fig.~2a.}
\label{fig:squid-harmonics}
\end{figure}

Figure 2a shows how, in a symmetrized configuration with balanced Josephson junctions,  the magnetic flux $\Phi_2$ can profoundly change the harmonic composition of the G-SQUID CPR with singularities at $\Phi_2 = \pm \Phi_0 / 4$ and $\Phi_2 = \pm \Phi_0 / 2$.
In order to gain more insight on this magnetic-flux control and to quantify the level  of harmonic ``distillation’' at the singularity points, we show in Fig.~\ref{fig:squid-harmonics}a the complete $\Phi_2$ dependence of the first four harmonics. The plotted amplitudes of these harmonics (colored  dots) are extracted from Figure ~\ref{fig:data}a by performing a Fourier transform of the measured $I_\mathrm{C}(\Phi_1)$ at every $\Phi_2$ value. The corresponding uncertainties are represented by $\pm 1 \sigma$-wide colored bands. These uncertainties are significant and visible only when the harmonic amplitudes are below $\sim 1$ nA, which is always the case for the fourth harmonic. 

The overlaid dashed lines represent the amplitudes of the first four harmonics calculated using the circuit model of Fig.~1a with model parameters obtained from the fitting of the data in Fig.~\ref{fig:data}a as previously discussed. The remarkable quantitative agreement over four orders of magnitude confirms the validity of our circuit model. 

At $\Phi_2 = \pm \Phi_0 / 2$, the first and third harmonics exhibit cusp-like dips where their amplitude is suppressed by two orders of magnitude, while the second and fourth harmonics simultaneously attain local maxima. In particular, the amplitude of the second harmonic at $\Phi_2 = \pm \Phi_0 / 2$ is almost identical to the one at $\Phi_2 = 0$. We can define a ``purity'' level of the $\sin(2\varphi)$ CPR as the ratio between the second harmonic amplitude, $A_{2\varphi}$, and the sum of the four harmonic amplitudes, $\Sigma_n A_{n\varphi}$. Its flux dependence is displayed in Fig.~\ref{fig:squid-harmonics}b. At $\Phi_2 = \pm \Phi_0 / 2$, we reach a second-harmonic purity of \SI{95.2(2.4)}\%, largely exceeding the state-of-the-art \cite{banszerus_voltage-controlled_2024,messelot_direct_2024}. This already high purity level could be further increased by additional circuit optimization. On the one hand, based on the circuit model of Fig.~1a, our measurement underestimates the  purity value due to the relatively small $I^\mathrm{W}_\mathrm{C}/I^\mathrm{G-SQUID}_\mathrm{C}$ ratio. We expect that a ten times larger $I^\mathrm{W}_\mathrm{C}/I^\mathrm{G-SQUID}_\mathrm{C}$ ratio would have resulted in a measured purity of \SI{96.3}{\percent}, much closer to the actual one. On the other hand, the intrinsic purity level could be increased by acting on other circuit parameters.

%Reference Julien Renard: \cite{messelot_direct_2024}

%Reference C. Marcus: \cite{banszerus_voltage-controlled_2024}

%Reducing the inductance of the G-SQUID arms from \SI{51}{\pH} to \SI{46}{\pH} or 
%Based on our circuit model, we find that this purity level is partially reduced by the relatively small $I^\mathrm{ref}_\mathrm{C}/I^\mathrm{G-SQUID}_\mathrm{C}$ ratio. A ten times larger ratio would have led to a measured purity of \SI{96.3}{\percent}.
%This ratio only limits the measurement of the purity, but not the purity itself, which, thanks to our model, is found to be \SI{96.3}{\percent}. 
%In the following we place ourselves in the $I^\mathrm{ref}_\mathrm{C}/I^\mathrm{G-SQUID}_\mathrm{C} \gg 1$ limit to give a broader point of view for future samples.
To illustrate that, we begin by noting that our fit of Fig.~2a reveals a slightly imperfect symmetry of the 
G-SQUID, quantified by a \SI{0.7}{\percent} discrepancy between the amplitudes of the first harmonics of $J_{N1}$ and $J_{N2}$ (see Supp.~S2). Reducing this discrepancy to less than \SI{0.14}{\percent} is in principle possible through a fine adjustment of the gate voltages. This would increase the weight of the even harmonics to more than \SI{99}{\percent}, with the fourth harmonic accounting for a few percent of the total weight. We note that both the second and the fourth harmonics contribute to parity protection since  they reflect the  simultaneous  transport of even numbers of Cooper pairs. As discussed before, the fourth harmonic is essentially absent in the CPR of the individual JoFETs, and it originates from the non-negligible inductance for G-SQUID arms. Based on our circuit model, we estimate that in a properly symmetrized G-SQUID, reducing the arm inductances $L_1$ and $L_2$ from \SI{51}{\pH} to \SI{46}{\pH} would largely suppress the fourth harmonic resulting in  $\sin(2\varphi)$ purity above \SI{99}{\percent}.

%First, the arm inductances of the G-SQUID induce an enhancement of the $\mathrm{4^{th}}$ harmonic  whose amplitude is not suppressed at $\Phi_2 = \Phi_0 / 2$. Our circuit model shows that a $\sin(2\varphi)$ purity of \SI{99}{\percent} could  be reached in G-SQUID with perfectly symmetrized Josephson juctions provided the arm inductances are lower than \SI{46}{\pH} (instead of \SI{51}{\pH} as in our G-SQUID).

%Second, the imperfect symmetry of the G-SQUID $\mathrm{1^{st}}$ and $\mathrm{3^{rd}}$ harmonics prevents their complete cancellation by destructive interference. It is therefore necessary to maintain these asymmetries below \SI{0.14}{\percent} (instead of \SI{0.7}{\percent} as in our G-SQUID) in order to achieve a $\sin(2\varphi)$ purity of \SI{99}{\percent}. 

%\begin{figure}
%\vspace{0.5cm}
%\includegraphics[width=0.45\textwidth]{Fig/fig3_V6.pdf}
%\caption{\textbf{Flux modulation of the harmonics amplitude}  \textbf{a}, Amplitudes of the 4 first harmonics contributions to the $\mathrm{G-SQUID}$ CPR as function of $\Phi_2$ re-scaled in unit of $\Phi_0$. The $\mathrm{1^{st}}$ harmonic vanishes at half flux quantum while the $\mathrm{2^{nd}}$ one vanishes at $\Phi_0/4$. \textbf{b}, The measured purity of the $\sin(2\varphi)$ Josephson element, defined as the contribution of the $\mathrm{2^{nd}}$ harmonic to the total $\mathrm{I_\mathrm{C}}$ amplitude, is then maximized at $\Phi_0/2$ and reaches \SI{95.2(2.4)}{\percent}. In black dashed line, we show the value of the G-SQUID CPRs Fourier transform from our circuit model.
%}
%\label{fig:squid-harmonics}
%\end{figure}

Finally, at $\Phi_2 = \pm \Phi_0 / 4$ the contribution of the second harmonic goes down to \SI{2.6(0.1)}{\percent}. This suppression of the second harmonic could be further pushed below \SI{1}{\percent} by changing the gate configuration (see Supp.~S5).

\section{Discussion and Conclusions}

Our study provides important insight to devise practical realizations of parity-protected qubit based on the demonstrated $\sin(2\varphi)$ Josephson element \cite{schrade_protected_2022, larsen_parity-protected_2020}. In principle, parity protection enhances the qubit lifetime at the cost of rendering the qubit resilient to external control pulses. Therefore, qubit operation would require temporarily exiting the protected regime, e.g. by controlling the ratio between the second and first harmonics. We show here that this ratio can be tuned by a magnetic flux or a gate voltage. 
A flux shift $\delta\Phi_2 = 0.025\Phi_0$ from $\Phi_2 = \Phi_0/2$ lower the second-harmonic content from \SI{95}{\percent} to \SI{60}{\percent}. 
In Supp.~S4 we show that, starting from a symmetric biasing of the G-SQUID, a gate voltage shift $\delta V_\mathrm{G}=\SI{20}{\milli\V}$ in one of the two JoFETs lowers the $\sin(2\varphi)$ purity by about \SI{25}{\percent}. Such flux and gate-voltage shifts are experimentally accessible with electrical control pulses on the typical time scale ($\sim$ 10 ns) of single-qubit operations. 

In conclusions, we have reported an experimental realization of a $\sin(2\varphi)$ Josephson element leveraging the intrinsic multi-harmonicity and gate tunability of SiGe-based JoFETs. The CPRs of these JoFETs can be accurately described by the sum of only three harmonics albeit higher harmonics are measured due to the relatively modest ratio  between  $I^\mathrm{ref}_\mathrm{C}$ and $I^\mathrm{W}_\mathrm{C}$.
The almost  complete suppression of the odd harmonics at half a flux quantum leaves a $\sin(2\varphi)$ CPR with a remarkably high purity level exceeding \SI{95}{\percent}. We argued that even higher values beyond \SI{99}{\percent} could be reached through further circuit optimization.    
All data analysis was based on a relatively simple but  accurate circuit model taking into account the non-sinusoidal CPRs of the JoFETs as well as the inductances of the superconducting arms.

\begin{acknowledgments}
This work has been supported by the ANR project SUNISIDEuP (ANR-19-CE47-0010), the PEPR ROBUSTSUPERQ and PRESQUILE (ANR-22-PETQ-0003), the ERC starting grant LONGSPIN (Horizon 2020 - 759388) and the Grenoble LaBEX LANEF. We thank the PTA (CEA-Grenoble) for the nanofabrication. We thank J. Renard and S. Messelot for discussions.
\end{acknowledgments}

\bibliography{biblio}

\begin{thebibliography}{48}
\bibitem [{foo()}]{footnote}%
  \BibitemOpen
  \bibfield  {title} {\bibinfo {title} {From our model we conclude that the 4\textsuperscript{th} harmonic is completely due to branch inductances},\ }\href@noop {} {\ }\BibitemShut {NoStop}%
\end{thebibliography}

%%%%%%%%%% Merge with supplemental materials %%%%%%%%%%
\widetext
\clearpage

\begin{center}
\textbf{\large Gate and flux tunable sin(2$\varphi$) Josephson element in proximitized junctions}
\end{center}

\begin{center}
\textbf{\large Supplemental Materials}
\end{center}
%%%%%%%%%% Merge with supplemental materials %%%%%%%%%%
%%%%%%%%%% Prefix a "S" to all equations, figures, tables and reset the counter %%%%%%%%%%
\setcounter{equation}{0}
\setcounter{figure}{0}
\setcounter{table}{0}
\setcounter{section}{0}
\setcounter{page}{1}
\makeatletter
\renewcommand{\theequation}{S\arabic{equation}}
\renewcommand{\thefigure}{S\arabic{figure}}
\renewcommand{\thetable}{S\arabic{table}}
\renewcommand{\thesection}{S-\Roman{section}}
%%%%%%%%%% Prefix a "S" to all equations, figures, tables and reset the counter %%%%%%%%%%

\section{JoFET geometry and arm inductances}

\begin{figure}[h]
\centering
\includegraphics[width=0.65\textwidth]{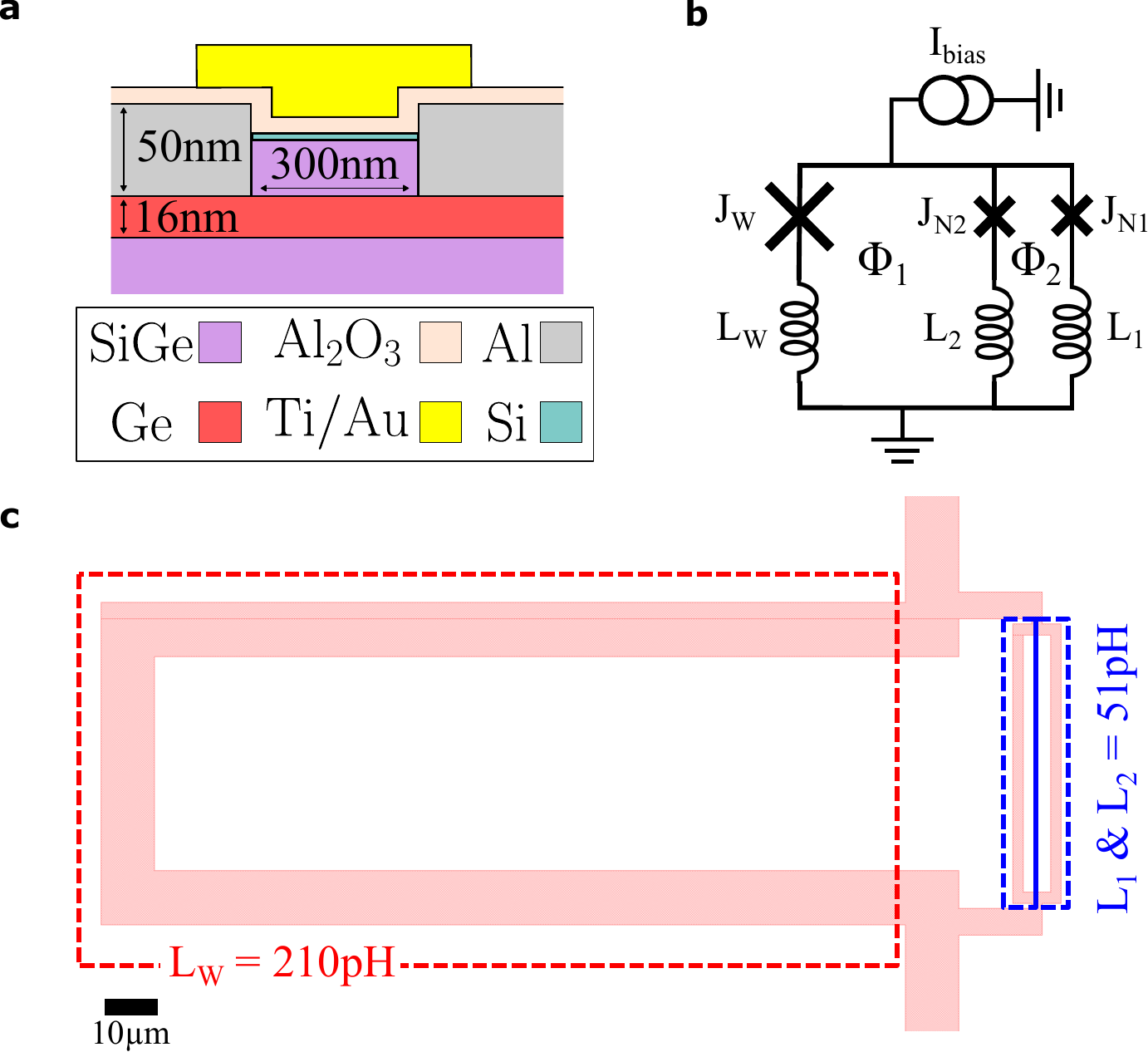}
\caption{\textbf{JoFET geometry and equivalent circuit of the double SQUID including arm inductances.} \textbf{a}, Cross-sectional sketch of a JoFET in which the Ge quantum well is contacted by two superconducting aluminum leads. The Ti/Au top gate allows tuning the hole carrier density in the channel. \textbf{b}, $J_{\mathrm N1}$, $J_{\mathrm N2}$ and $J_\mathrm W$ are the JoFETs, $L_\mathrm 1$, $L_\mathrm 2$ and $L_\mathrm 3$ are the inductances of the three aluminum arms. The device is current biased by $I_{bias}$. \textbf{c}, Design of the 3 aluminum arms with their respective inductances calculated by finite element simulation using the Sonnet software.}
\end{figure}

The inductances $L_\mathrm 1$, $L_\mathrm 2$ and $L_\mathrm W$ have a geometric and a kinetic contribution $L_\mathrm{geo}$ and $L_\mathrm{kin}$. The kinetic inductance per square $L_\mathrm{kin}^\mathrm{S}$ is estimated from the aluminum normal state sheet resistance $R_\mathrm{S} = \SI{0.87}{\ohm}$ and its superconducting critical temperature $T_\mathrm{C}=\SI{1.5}{\kelvin}$ (both measured on the same chip) \cite{mattis_theory_1958, tinkham_introduction_2015}:
\begin{equation}
    L_\mathrm{kin}^S \approx \frac{\hbar}{\pi} \frac{R_S}{1.76 k_B T_C}
\end{equation}
where $k_B$ is the Boltzmann constant. We find $L_\mathrm{kin}^S = \SI{0.8}{\pico\henry\per\sq}$. To estimate the total loop inductance  $L = L_\mathrm{kin} + L_\mathrm{geo}$, we use a finite element simulation performed in Sonnet and find $L_\mathrm{1} \& L_\mathrm{2}=\SI{51}{\pico\henry}$ and $L_\mathrm{W}=\SI{210}{\pico\henry}$.

The inductance of the G-SQUID loop and the presence of higher harmonics in the junction CPR both lead to skewness in the $I_\mathrm{C}$ versus $\Phi$ data.

\newpage

\section{$I_C$ VS $\Phi$ model with inductance}

\subsection{Theoretical model}

To model the device behaviour we approximate the CPR of junction X (with $X \in \{W,N1,N2\}$) by a sum of $n$ sinusoidal harmonics:
\begin{equation}
    I_X(\varphi) = \sum_{n} (-1)^{n-1} H_n^X \sin(n\varphi)
\end{equation}
where $H_n^X$ is the magnitude of the $n^{th}$ harmonic. We note $\varphi_W$, $\varphi_{N1}$ and $\varphi_{N2}$ the phases across each junction. Thus the overall current, $I_{tot}$, flowing through the device satisfies the following equation:
\begin{equation}
    I_{tot}(\varphi_{W},\varphi_{N1},\varphi_{N2}) = I_{W}(\varphi_W) + I_{N1}(\varphi_{N1}) + I_{N2}(\varphi_{N2})
\end{equation}
Each arm of the double SQUID device contains an inductance (denoted L1, L2, and L3) and the two loops are flux biased by fluxes $\Phi_1$ and $\Phi_2$, as depicted in Fig.~S1. Consequently, the fluxoid quantification yields the following relations: 
\begin{equation}
    \varphi_{N2} = \varphi_{W} + \frac{2\pi}{\Phi_0} \bigl[ L_W I_{W}\left(\varphi_W\right) - L_2 I_{N2}\left(\varphi_{N2}\right) + \Phi_1 \bigr]
\end{equation}
\begin{equation}
    \varphi_{N1} = \varphi_{N2} + \frac{2\pi}{\Phi_0} \bigl[ L_2 I_{N2}\left(\varphi_{N2}\right) - L_1 I_{N1}\left(\varphi_{N1}\right) + \Phi_2 \bigr]
\end{equation}
In order to ascertain the critical current of the device, it is necessary to identify the maximum supercurrent $I_{tot}$ for which there exists a phase configuration $\{\varphi_{W}, \varphi_{N1}, \varphi_{N2}\}$ that satisfies S3,4 and 5. To this end, we must incorporate this additional information into our equation system:
\begin{equation}
    \frac{\partial I_{W}}{\partial \varphi_{W}} = 0 \quad\quad
    \frac{\partial I_{N1}}{\partial \varphi_{N1}} = 0 \quad\quad
    \frac{\partial I_{N2}}{\partial \varphi_{N2}} = 0
\end{equation}

Consequently, the equation system S3,4,5, and 6 can be solved numerically to determine the critical current of the device at any flux configuration $\Phi_1$, $\Phi_2$.

\subsection{Fit to the model}

The aforementioned model is employed to fit the G-SQUID flux response depicted in Fig.~2a. The resulting harmonic composition of the fit is shown in Fig.~S2a. The arm inductances are fixed to the values estimated in S1, and the harmonic composition resulting from the fit is reported in Table S1. For purposes of comparison, the same model is also shown in Fig.~S2b, with the harmonic amplitudes fixed to the values measured in Fig. 1e,f.

\begin{table}[h]
\begin{tabular}{c|c|c|c|c|c|c|c|c|c|c|c|c|c|c|c}
Parameter           & $L_1$ & $L_2$ & $L_W$ & $H^W_1$ & $H^W_2$ & $H^W_3$ & $H^W_4$ & $H^{N1}_1$ & $H^{N1}_2$ & $H^{N1}_3$ & $H^{N1}_4$ & $H^{N2}_1$ & $H^{N2}_2$ & $H^{N2}_3$ & $H^{N2}_4$ \\ \hline
Value               & 51pH & 51pH  & 210pH  & 3.04µA  & 174nA   & 116nA   & 0nA     & 282.6nA      & 31.8nA       & 1.6nA        & 0nA        & 284.6nA      & 32.0nA       & 1.6nA        & 0nA        \\ \hline
Fit                 & fixed & fixed & fixed & free    & free    & free    & free    & free       & free       & free       & free       & free       & free       & free       & free       \\ \hline
Error & N/A  & N/A  & N/A  & 0.03\%  & 3.18\%  & 1.14\%  & N/A    & 0.02\%     & 0.18\%     & 3.51\%     & N/A       & 0.02\%     & 0.18\%     & 3.48\%     & N/A      
\end{tabular}
\caption{\textbf{Fitting parameters considering the full model and the inductances from S1} The result is plotted in Fig.~S2a.}
\end{table}

\begin{table}[h]
\begin{tabular}{c|c|c|c|c|c|c|c|c|c|c|c|c}
Parameter & $L_1$ & $L_2$ & $L_W$ & $H^W_1$ & $H^{N1}_1$ & $H^{N1}_2$ & $H^{N1}_3$ & $H^{N1}_4$ & $H^{N2}_1$ & $H^{N2}_2$ & $H^{N2}_3$ & $H^{N2}_4$ \\ \hline
value     & 51pH & 51pH  & 210pH  & 2.9µA   & 303.2nA    & 41nA       & 5.6nA      & 2.12nA     & 302.9nA    & 41.46nA    & 5.5nA      & 2.03nA     \\ \hline
fit       & fixed & fixed & fixed & fixed   & fixed      & fixed      & fixed      & fixed      & fixed      & fixed      & fixed      & fixed     
\end{tabular}
\caption{\textbf{Full model with inductances from S1 and harmonics amplitudes from measurements Fig.~1e,f.} The result is plotted in Fig.~S2b.}
\end{table}

\begin{figure}[h]
\centering
\includegraphics[width=1\textwidth]{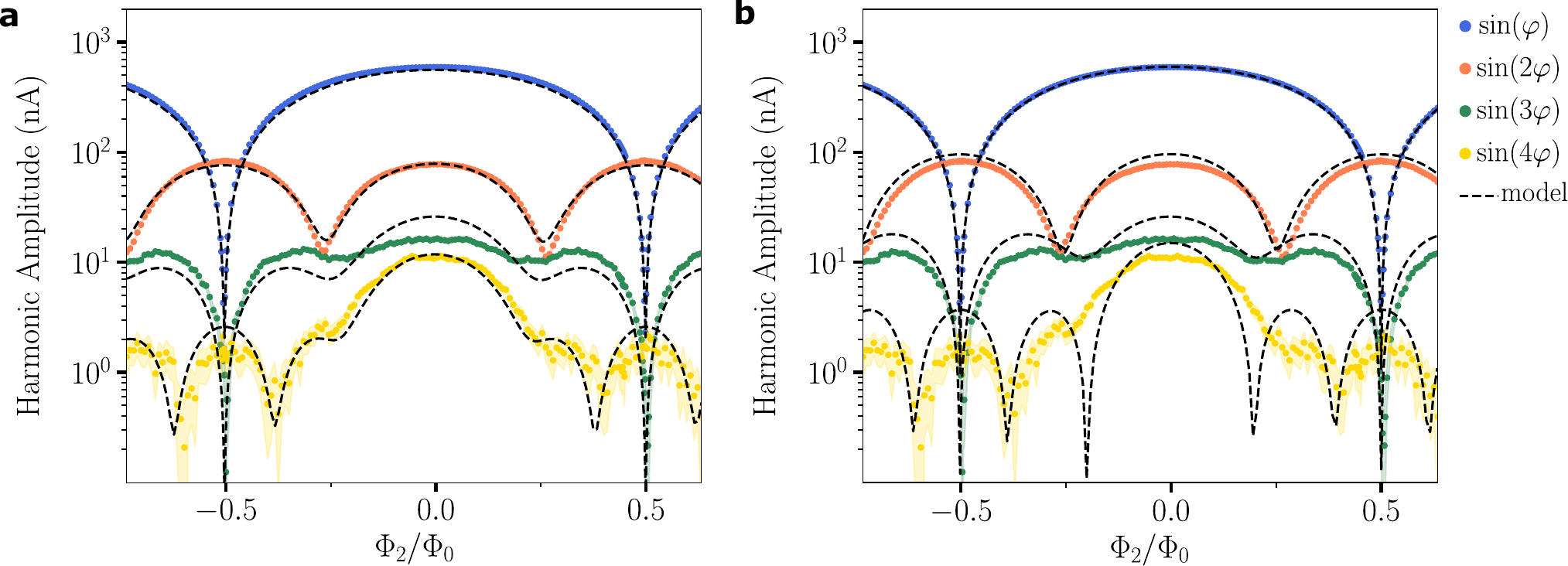}
\caption{\textbf{Model including arm inductances.}  \textbf{a}, Model including the inductances from S1 while the JJs harmonic contents (reported in table S1) are given by the fit. \textbf{b}, Model including the inductances from S1 while the JJs harmonic contents are set to those measured in Fig.~1e,f (reported in Table S2).}
\end{figure}

\begin{figure}[h]
\centering
\includegraphics[width=0.8\textwidth]{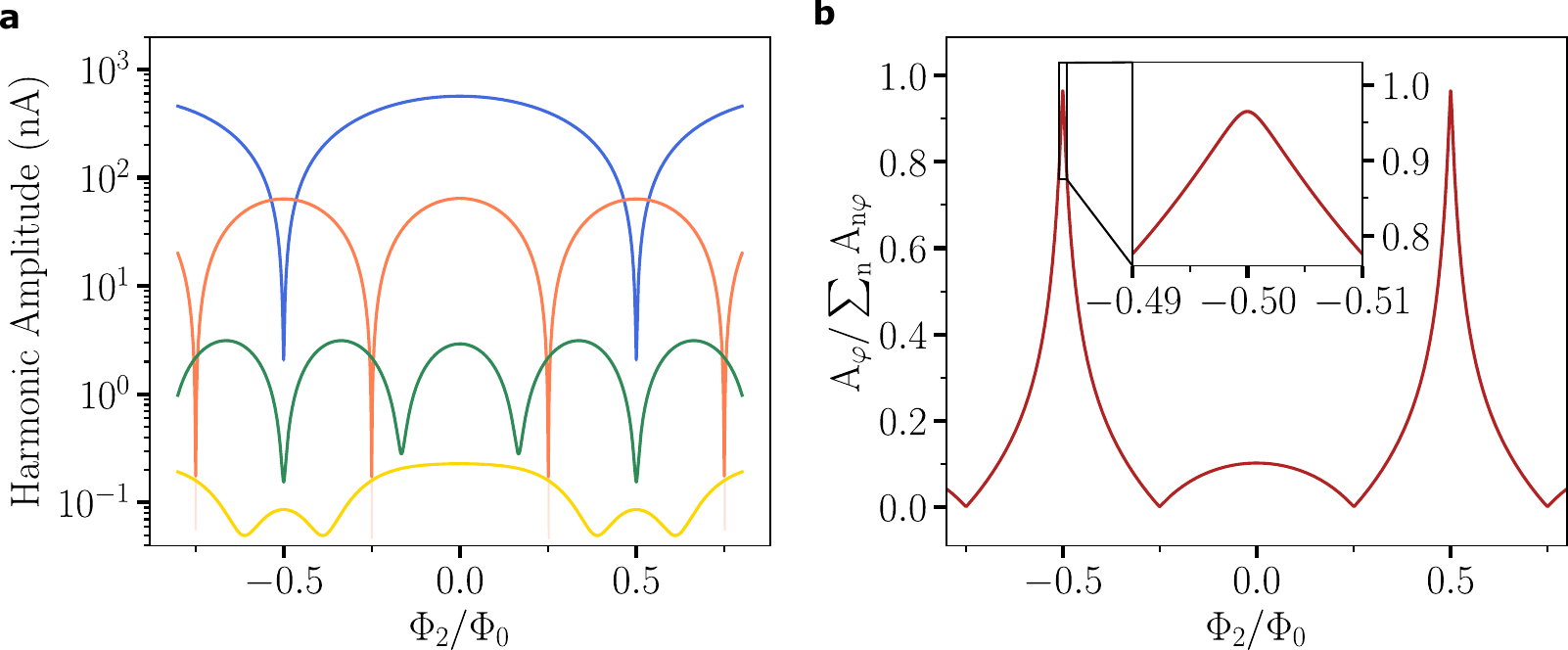}
\caption{\textbf{Theoretical G-SQUID CPR in the absence of arm inductances.} The junctions CPR compositions are obtained from the fit shown in Fig.~S2a and listed in Table S2. The confidence interval provided by the fit, along with the correlation matrix, is employed to compute the confidence interval of the G-SQUID CPR composition, which is found to be negligible. \textbf{a}, Resulting G-SQUID harmonics amplitude as a function of flux threading $\Phi_2$.  \textbf{b}, Associated $\sin(2\varphi)$ purity.}
\end{figure}

\begin{figure}[h]
\centering
\includegraphics[width=0.8\textwidth]{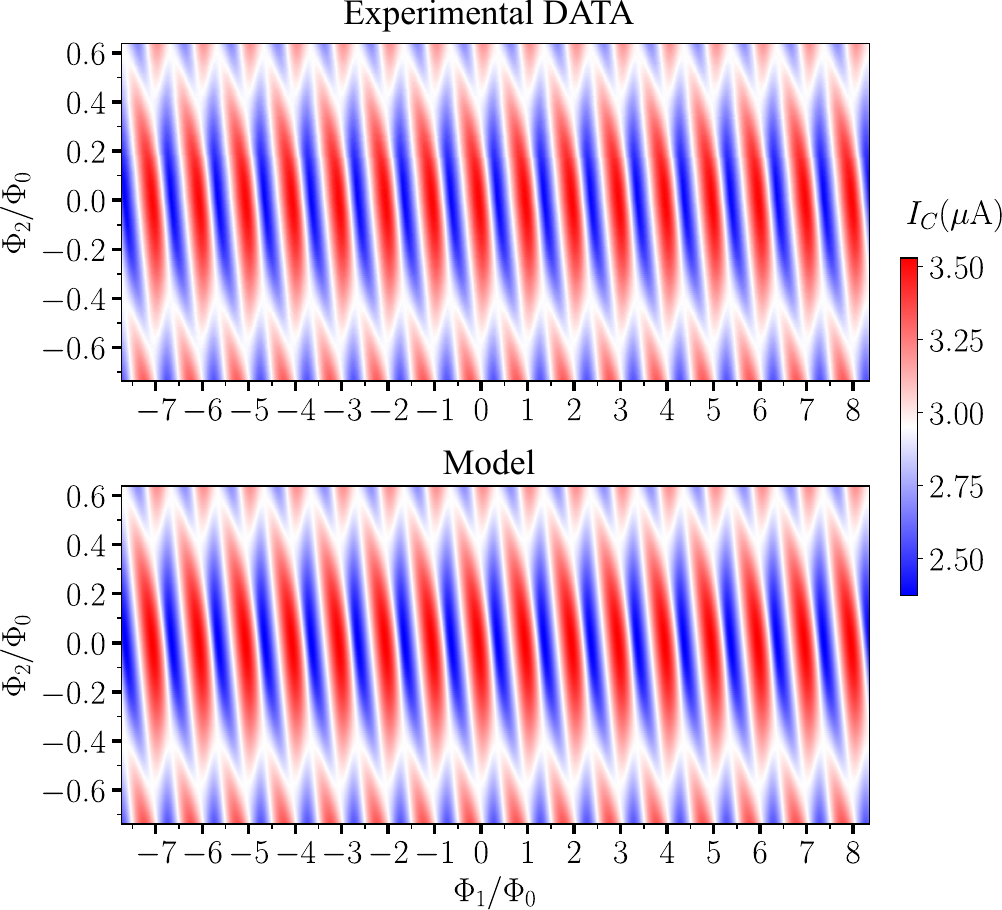}
\caption{\textbf{Full data set fit.} \textbf{Top panel}, Full $\Phi_1$ VS $\Phi_2$ data set (i.e. zoom out of Fig.~2a).  \textbf{Lower panel}, Result from the fit using the full model described above and the fit parameters shown in Tab.~S1.}
\end{figure}

It can be observed that the inclusion of arm inductances in the model, while maintaining the harmonic content as measured in Fig.~1e,f, is insufficient to accurately reproduce the data. However, by fitting the data with the model, it is possible to identify a set of parameters (i.e., the harmonic content of each junction) that closely aligns with the experimental results.

Furthermore, the fit has enabled the determination of the "real" harmonic content of each junction. This allows for the estimation of the theoretical harmonic content of the G-SQUID in the absence of arm inductance. With a simple SQUID model without inductance, the G-SQUID harmonic content is obtained, as shown in Fig.~S3. This theoretical harmonic content is comparable to the measured one, with the exception that the fourth harmonic is considerably smaller. Consequently, we can conclude that in this ideal situation, the $\sin(2\varphi)$ purity, which reaches \SI{96.5}{\percent}, cannot be limited by the presence of a fourth harmonic that is not suppressed at $\Phi_2=\Phi_0/2$. Therefore, the remaining limiting parameter to reach the perfect $\sin(2\varphi)$ regime is the G-SQUID symmetry, in particular the symmetry of the first and third harmonics.

It is important to note that in order to accurately measure very high $\sin(2\varphi)$ purity using a double SQUID device, it is essential to consider the ratio between the reference junction and the G-SQUID critical currents. This ratio must be sufficiently large to prevent any artificial limitation on the figure of merit by enhancing higher order harmonics. This limitation can be illustrated by the G-SQUID CPR measured at $\Phi_2=0$ and shown in Fig.~2b. The CPR exhibits sharp corners in the minima and smooth maxima, thereby breaking the time-reversal symmetry. This is an artefact resulting from the mixing of the reference junction and G-SQUID CPRs, which is made possible by a too small ratio $I^{ref}_C/I^{SQUID}_C$ when $\Phi_2=0$. For further details, please refer to the detailed analysis presented in \cite{babich_limitations_2023}. However, this artifact disappears around the point of interest, $\Phi_2 = \Phi_0/2$, since the ratio $I^{ref}_C/I^{SQUID}_C$ is greater than 20 for this flux.  
\newpage

\section{Harmonics phases}

In the main text, the harmonics of the G-SQUID CPR are extracted by Fourier transformation, and their respective absolute amplitudes are shown in Fig.3. Here, we show in Fig.~S4 the phase of each harmonic as a function of the flux biasing the G-SQUID. The behaviors of these harmonics are well reproduced by the model described in Supp.~S2 and shown with black dotted lines. The flux $\Phi_2$ threading the G-SQUID dictates the phase difference across the G-SQUID. Consequently, the average slope of the $n^{th}$ harmonic phase with respect to $\Phi_2$ is:
\begin{equation}
    \frac{\partial\varphi_n}{\partial\Phi_2} = n \pi \Phi_0
\end{equation}
Furthermore, each time the $n^{th}$ harmonic is canceled (i.e. dips in Fig.3), its phase experience a $\pi$ shift.

\begin{figure}[h]
\centering
\includegraphics[width=0.5\textwidth]{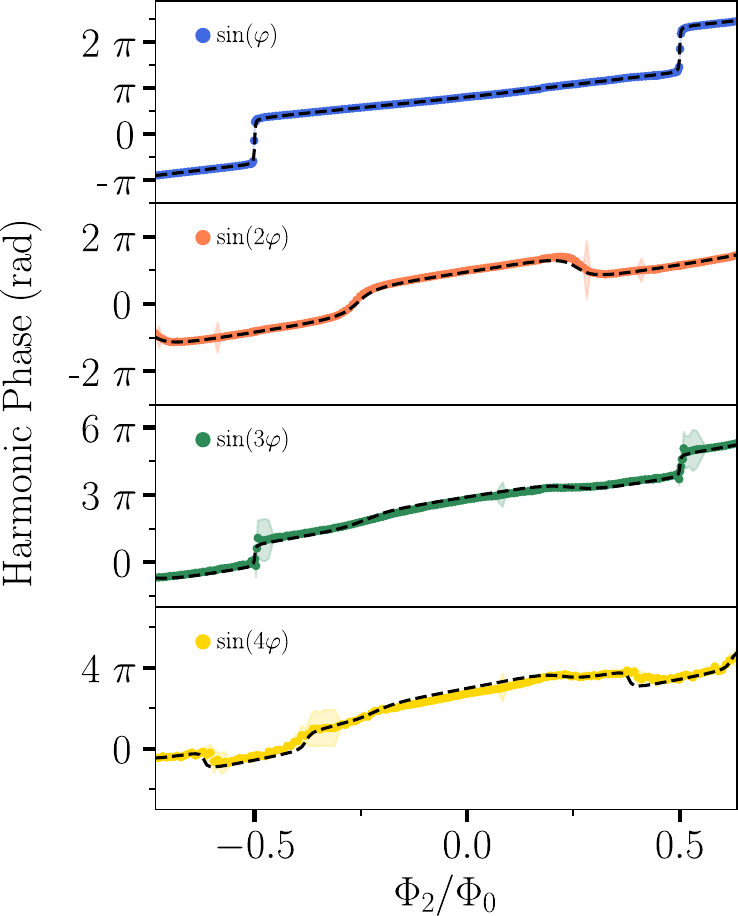}
\caption{\textbf{Flux dependence of harmonics phases.} The phase of each harmonic of the G-SQUID CPR extracted by Fourier transformation. The aforementioned phases are presented as a function of the flux $\Phi_2$ threading the G-SQUID. The black dotted lines represent the phases extracted by the same methodology from the model discussed in Supp.~S2.}
\end{figure}

\newpage

\section{Gate tuning of G-SQUID symmetry}
We demonstrated the flux dependence of the harmonic content of the symmetric G-SQUID CPR. An alternative method for exiting the "$\sin(2\varphi)$" regime is to detune the G-SQUID from symmetry. In this experiment, the flux through the SQUID was set to $\Phi_2=\Phi_0/2$.

In the initial stage of the experiment, the $J_{N2}$ gate voltage was set to $V_G^{N2}=\SI{-1.5}{\V}$ in order to maintain a full accumulation, while the $J_{N1}$ gate voltage was varied as illustrated in Fig.~S5a. In Fig.~S5b, we identify the reduction of the first harmonic around $V_G^{N1}=\SI{0.7}{\V}$. Consequently, this gate configuration corresponds to the symmetric G-SQUID regime. Figure S5c illustrates the phase evolution of each harmonic of the G-SQUID CPR. It can be observed that even harmonics maintain a constant phase, while a $\pi$ shift occurs in odd phases around the symmetric regime point (see Supp.~S4).

In a second instance, we examine a configuration where the G-SQUID symmetry is more susceptible to gate voltage. The gate voltage of $J_{N2}$ is fixed at a value of $V_G^{N2}=\SI{-0.2}{\V}$, while the gate voltage of $J_{N1}$  is ramped between \SI{-0.3}{\V} and \SI{0.25}{\V}, as indicated in Fig.~S5d. As illustrated in Fig.~S5e, the harmonic dependence of the gate is more pronounced in this configuration than in the previous one. In such a configuration, we demonstrate that a gate voltage shift of \SI{20}{\mV} on $V_G^{N1}$ permits the achievement of a purity shift from \SI{91}{\percent} to \SI{67}{\percent}.

\begin{figure}
\centering
\includegraphics[width=0.95\textwidth]{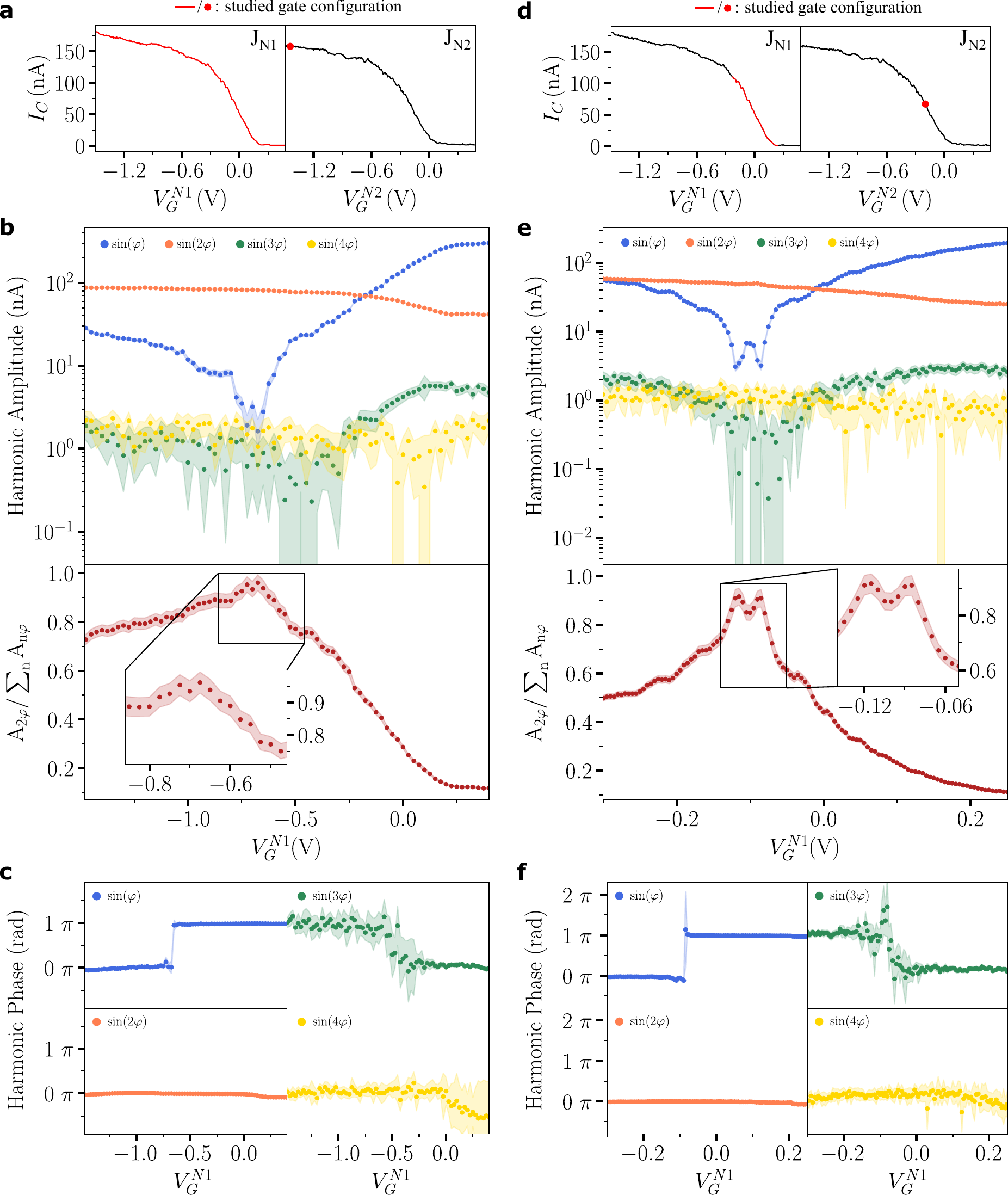}
\caption{\textbf{Gate modulation of the G-SQUID CPR.} The left column depicts the gate modulation in the high accumulation regime, while the right column illustrates it in the lower accumulation. \textbf{a[d]}, Gate configurations explored: The $J_{N1}$ gate voltage is swept along the red line (left panel) while the $J_{N2}$ gate voltage is maintained at the value represented by the red point (right panel). \textbf{b[e]}, amplitudes of the four first harmonics (top panel) and the $\sin(2\varphi)$ purity (bottom panel) with respect to the $J_{N1}$ gate voltage. \textbf{c[f]}, phases of the four first harmonics with respect to the $J_{N1}$ gate voltage.}
\end{figure}

\section{Second harmonic canceling}

In the main text, we focused on the cancellation of the first harmonic in order to engineer a CPR dominated by the second harmonic. To do so the gate configuration was chosen to equalize the first harmonics of the two junctions of the G-SQUID because the more symmetric they are, the more they are cancelled out at $\Phi_2=\Phi_0/2$. In this section, we investigate the suppression of the second harmonic and choose a suitable gate configuration: $V_G^{N1}=\SI{-0.15}{\V}$ and $V_G^{N2}=\SI{-1.5}{\V}$. Figure S6a shows the resulting CPR as function of $\Phi_2$. Line cuts and Fourier transforms (Fig.~S6b,c,d,e,f,g) highlight the CPR at $\Phi_2=0$, $\Phi_2=\Phi_0/4$ and $\Phi_2=\Phi_0/2$. It is observed that the second harmonic is completely suppressed at $\Phi_0/4$, but that the first harmonic is no longer completely canceled at $\Phi_0/2$. This is due to the fact that we cannot reach perfect symmetry in the G-SQUID for both first and second harmonics.

Fig.~S7 shows the amplitudes and phases of each harmonic with respect to the flux $\Phi_2$ threading the G-SQUID. The lower panel of Fig.~S7a shows the first harmonic purity defined as ratio between the first harmonic amplitude and the total supercurrent amplitude. The maximum value reported is \SI{97.3(0.6)}{\percent}. It is noteworthy that this range of $\sin(\varphi)$ purity is comparable to state of the art aluminum tunnel junctions as recently reported by Willsch et al. \cite{willsch_observation_2024}.

%Using the same procedure as exposed before (Supp.~S2), we extract the harmonic content of each JoFET for this gate configuration (table S3). The resulting model is shown as black dotted lines in Fig.~S6 and S7.
%\begin{table}[h]
%\begin{tabular}{c|c|c|c|c|c|c|c|c|c|c|c|c|c|c|c}
%Parameter           & $L_1$ & $L_2$ & $L_3$ & $H^W_1$ & $H^W_2$ & $H^W_3$ & $H^W_4$ & $H^{N1}_1$ & $H^{N1}_2$ & $H^{N1}_3$ & $H^{N1}_4$ & $H^{N2}_1$ & $H^{N2}_2$ & $H^{N2}_3$ & $H^{N2}_4$ \\ \hline
%Value               & 210pH & 51pH  & 51pH  & 3.11µA  & 202nA   & 77nA   & 3nA     & 281.9nA      & 32.5nA       & 1.5nA        & 0nA        & 224.3nA      & 24.5nA       & 2.5nA        & 0nA        \\ \hline
%Fit                 & fixed & fixed & fixed & free    & free    & free    & free    & free       & free       & free       & free       & free       & free       & free       & free       %\\ \hline
%Error & N/A  & N/A  & N/A  & 0.03\%  & 3.18\%  & 1.14\%  & N/A    & 0.02\%     & 0.18\%     & 3.51\%     & N/A       & 0.02\%     & 0.18\%     & 3.48\%     & N/A      
%\end{tabular}
%\caption{\textbf{Fitting parameters considering the full model and the inductances from S1} The result is plotted in Fig.~S6 and S7.}
%\label{}
%\end{table}

\begin{figure}
\centering
\includegraphics[width=0.95\textwidth]{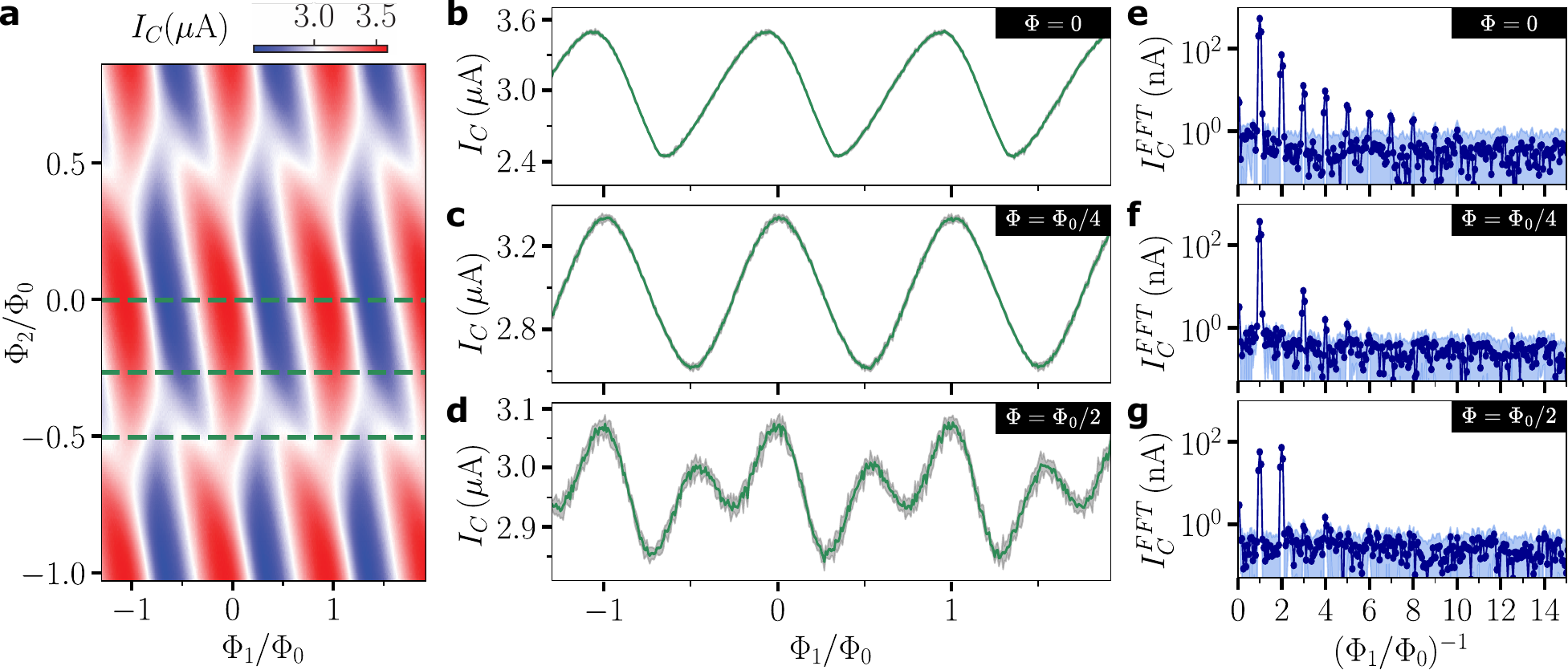}
\caption{\textbf{Second harmonic cancellation.} The G-SQUID is tuned to make the second harmonics symmetric and thus maximize their cancellation at $\Phi_0/4$: $V_G^{N1}=\SI{-0.15}{\V}$ and $V_G^{N2}=\SI{-1.5}{\V}$ \textbf{a}, Critical current as a function of both fluxes $\Phi_1$ and $\Phi_2$ revealing the G-SQUID CPR. \textbf{b[c][d]}, CPR at $\Phi_2=0$ [$\Phi_2=\Phi_0/4$] [$\Phi_2=\Phi_0/2$] and its Fourier decomposition \textbf{e[f][g]}.}
\end{figure}
\begin{figure}
\centering
\includegraphics[width=0.95\textwidth]{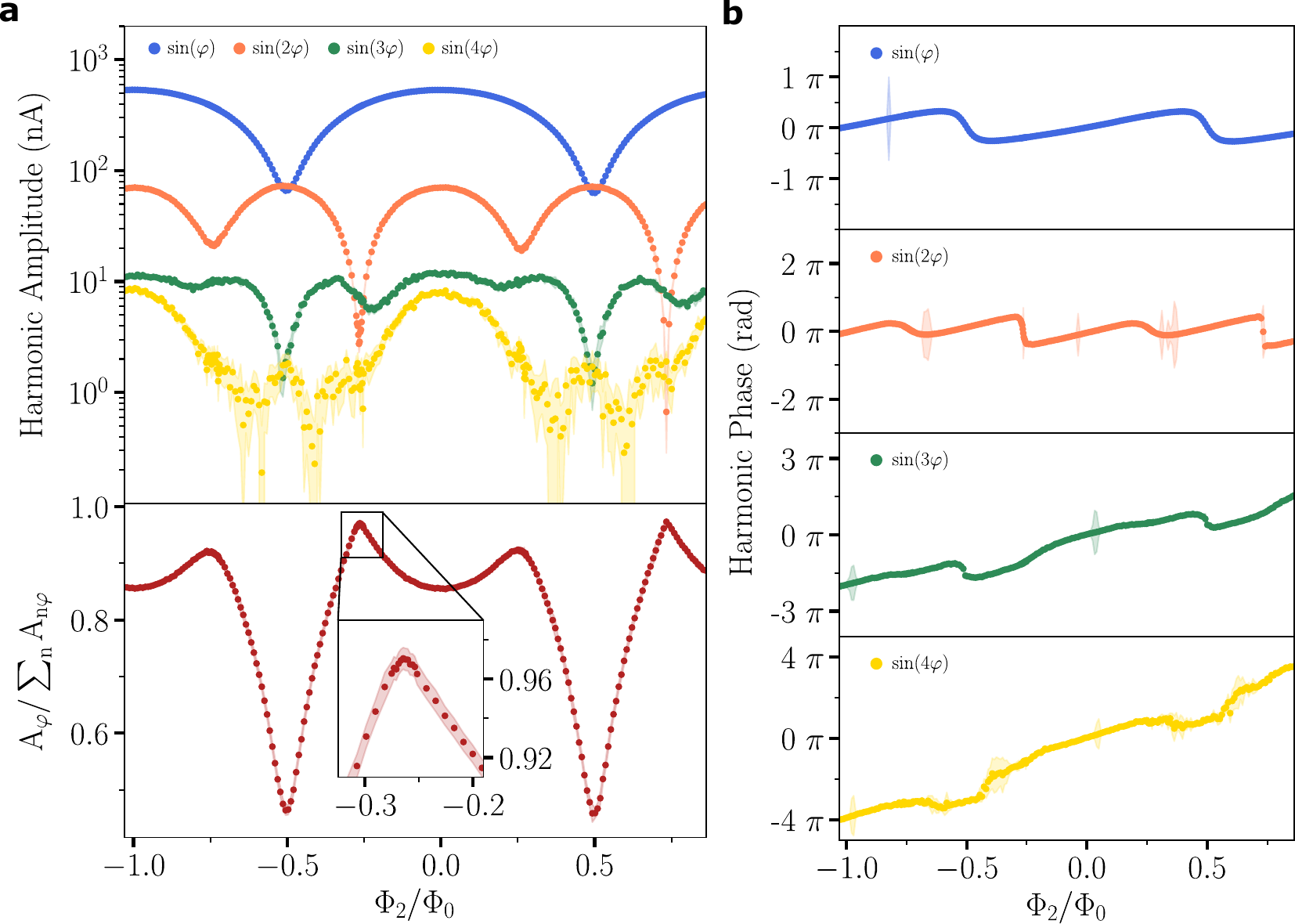}
\caption{\textbf{Nearly pure sinusoidal Josephson element.} The G-SQUID is tuned to make the second harmonics symmetric and thus maximize their cancellation at $\Phi_0/4$: $V_G^{N1}=\SI{-0.15}{\V}$ and $V_G^{N2}=\SI{-1.5}{\V}$ \textbf{a}, Upper panel: amplitude of each harmonic as function of the flux threading the G-SQUID loop. Lower panel: purity of the $\sin(\varphi)$ harmonic which reaches a value of \SI{97.3(0.6)}{\percent} when the second harmonic is cancelled out. \textbf{b}, Phase of each harmonic as function of $\Phi_2$.}
\end{figure}

\newpage

\section{Flux compensation}

Two local flux lines permit the independent flux biasing of flux loops. The two flux lines are current biased at \SI{4}{\kelvin} with a \SI{4.7}{\kilo\ohm} resistor as depicted in Fig.~S8a,d. In order to estimate the cross-talk between the two fluxes we conduct two experiments. First, the small G-SQUID is isolated by pinching off $J_W$ ($V_G^W=\SI{2}{\V}$) and the SQUID oscillations are measured by threading $\Phi_2$ with $V_1$ (Fig.~S8b) and with $V_2$ (Fig.~S8c). Secondly, $J_{N1}$ is pinched off to isolate an asymmetric SQUID comprising $J_W$ and $J_{N2}$. Again, SQUID oscillations are measured by threading the loop with $V_1$ (Fig.~S8e) and $V_2$ (Fig.~S8f).

By superimposing, Fig.~S8b with Fig.~S8c and Fig.~S8e with Fig.~S8f we deduce the following compensation matrix:
\[
    \left[ {\begin{array}{cc}
    V_{\Phi_1} \\
    V_{\Phi_2} \\
  \end{array} } \right] =
  \left[ {\begin{array}{cc}
    1 & 0.0935 \\
    0.182 & 1 \\
  \end{array} } \right]
  \left[ {\begin{array}{cc}
    V_{1} \\
    V_{2} \\
  \end{array} } \right]
\]

\begin{figure}[h]
\centering
\includegraphics[width=0.8\textwidth]{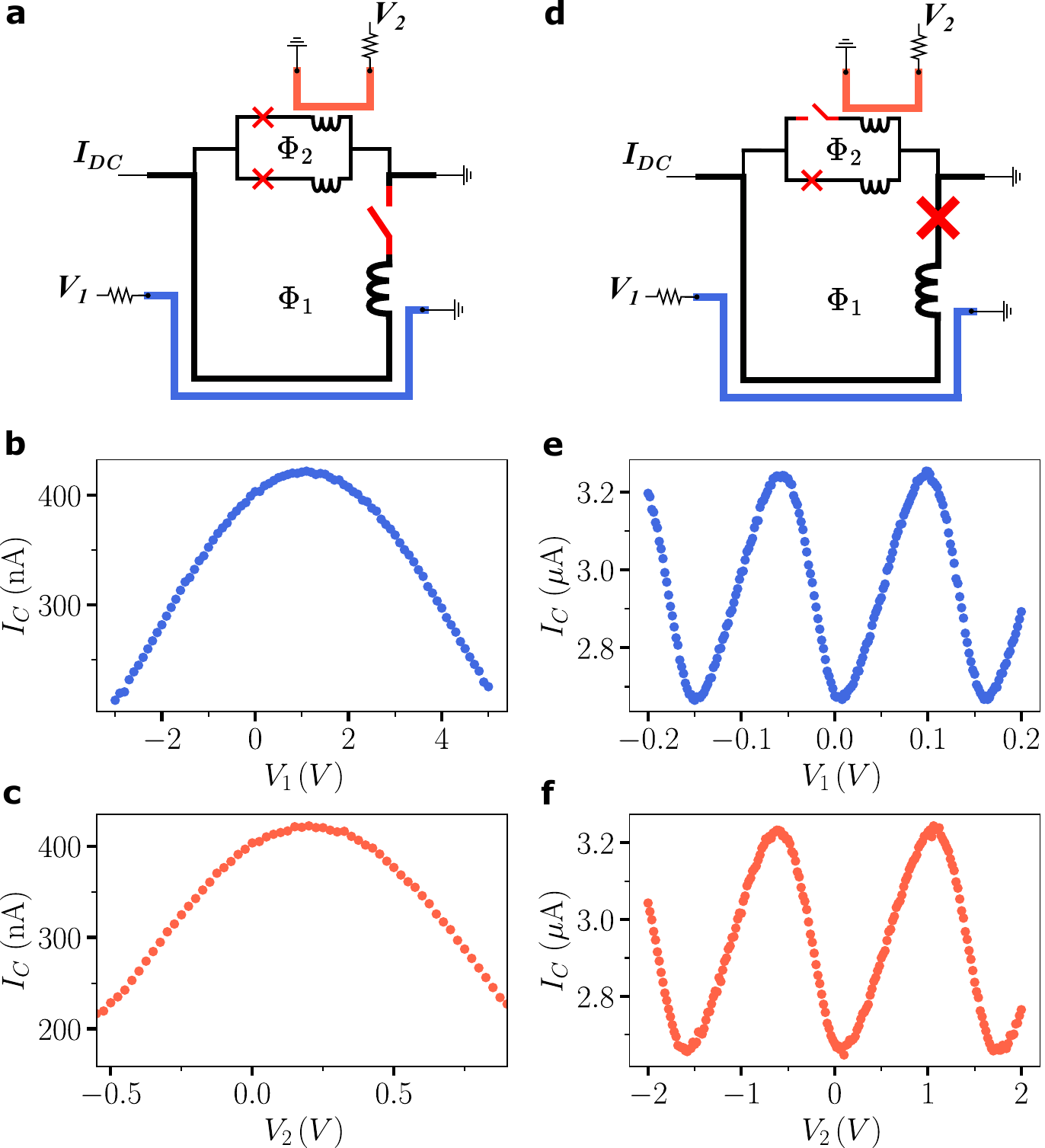}
\caption{\textbf{Flux cross-talk estimation.} \textbf{a}, The first configuration permits to explore the G-SQUID critical current oscillations as a function of the flux induced by each flux line \textbf{b} and \textbf{c}. \textbf{d}, The second configuration permits to measure the critical current oscillations of the SQUID formed by $J_W$ and $J_{N2}$ as a function of the flux induced by each flux line  \textbf{e} and \textbf{f}.}
\end{figure}

\newpage

\section{JoFETs threshold}

As illustrated in Fig.~1c and c of the main text, the two narrow JoFETs embedded in the G-SQUID, $J_{N1}$ and $J_{N2}$, exhibit a remarkably similar $I_C-V_G$ characteristic. However, a threshold voltage shift between the two is evident, as shown in Fig.~S9. The critical currents are extracted from the data shown in Fig.~1c and d and presented in the Fig.~S9 left panel. In the right panel, the critical current is plotted as a function of $V_G-V_{th}$, where $V_{th}$ is the threshold voltage. By manual adjustment, we find $V_{th}^{N1}=\SI{0.2}{\V}$ and $V_{th}^{N2}=\SI{0.05}{\V}$.

\begin{figure}[h]
\centering
\includegraphics[width=0.6\textwidth]{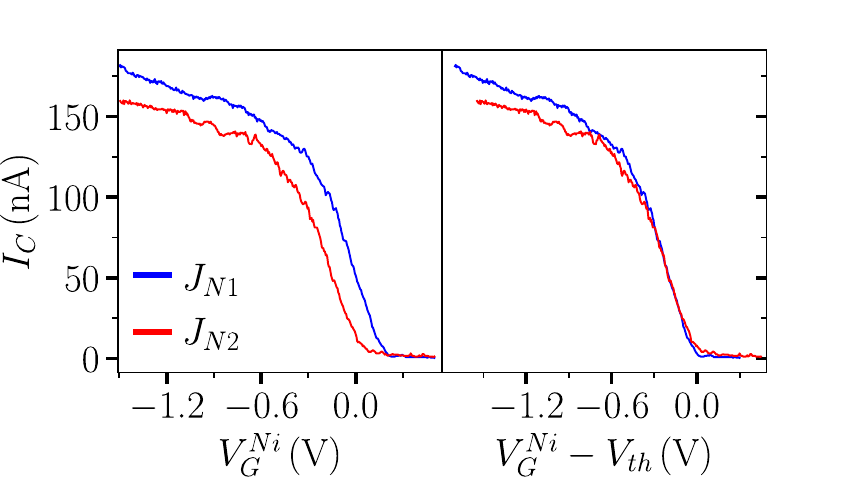}
\caption{\textbf{JoFETs threshold.} \textbf{Left}, $J_{N1}$ and $J_{N2}$ critical currents as a function of the gate voltage (extracted from the data shown in Fig.1c and d). \textbf{Right}, $I_C$ as a function of $V_G-V_{th}$ where $V_{th}$ is the threshold voltage. We find $V_{th}^{N1}=\SI{0.2}{\V}$ and $V_{th}^{N2}=\SI{0.05}{\V}$.}
\end{figure}

%\newpage
%\bibliography{biblio}

\end{document}